\documentstyle[12pt]{article}
\input epsf.tex

\newcommand{\NC}{\newcommand}
\NC{\beq}{\begin{equation}}
\NC{\eeq}{\end{equation}}
\NC{\beqa}{\begin{eqnarray}}
\NC{\eeqa}{\end{eqnarray}}

\begin{document}

\title{Generalised cosmology of codimension-two braneworlds}

\author{J\'er\'emie Vinet\\Physics Department, McGill University, Montr\'eal, Qu\'ebec, Canada H3A 2T8\\vinetj@physics.mcgill.ca}
\maketitle

\begin{abstract}
{It has recently been argued that codimension-two 
braneworlds offer a promising line of attack on the cosmological 
constant problem, since in such models the Hubble rate is not 
directly related to the brane tension.  We point out challenges 
to building more general models where the brane content is not 
restricted to pure tension.  In order to address these challenges, 
we construct a thick brane model which we linearize around a 
well known static solution.  We show that the model's cosmology 
does reduce to standard FRW behaviour, but find no hint of a 
self-tuning mechanism which might help solve the cosmological 
constant problem whithin the context of non-supersymmetric 
Einstein gravity.  
}
\end{abstract}

\section{Introduction}

It is now widely accepted by the physics community that the most 
recent high-precision cosmological data implies that the Universe 
is currently experiencing a state of accelerated expansion 
\cite{SN1a,WMAP}.  Such observations are consistent with the presence 
of a non-zero, but surprisingly small, cosmological constant.  It has 
thus become a pressing challenge for theoretical physicists to come up 
with an explanation for the presence and size of such a contribution to 
the Universe's energy content.  

Lately, reviving ideas that had been developped some twenty years ago 
\cite{Rubakov:1983bz,Randjbar-Daemi:1985wg,Gibbons:1986wg}, a promising 
avenue for tackling this problem has surfaced in the context of braneworld 
scenarios \cite{ADD,RSI,RSII}, more precisely in codimension-two braneworlds 
\cite{Chen:2000at,Carroll,Navarro1,cdgv}.  The key observation is that in such models, 
the four dimensional expansion rate is not directly related to the brane 
tension, but is rather a function of bulk parameters.  This feature is 
intimately related to the fact that codimension-two objects induce a 
conical singularity, but do not otherwise curve the surrounding space.  
  
While the cosmology of codimension-one branes have been well studied 
\cite{BDL,CGS}, the same can not be said of codimension-two branes, for 
which research has so far centered almost exclusively on maximally symmetric 
solutions with pure tension branes.  The reason for this, as we will show, is 
that for more general types of brane energy content, the metric becomes 
singular at the position of the brane(s), so that some sort of regularisation 
scheme is needed in order to study the behaviour of these models in a more 
generalised setting.  

We present in this talk summary of work done in \cite{vc}, where we have 
constructed a thick braneworld model in order to answer the following 
questions:\begin{itemize}
\item does the cosmology of codimension-two braneworlds follow the expected 
FRW behaviour?
\item how does the deficit angle react to a sudden change in the brane tension, as 
happens e.g. during a phase transition?
\end{itemize}

\section{Codimension-two branes}

The feature which makes codimension-two branes an attractive possibility for 
addressing the cosmological constant problem is the fact that the brane does 
not curve the surrounding internal space, apart from inducing a conical defect 
proportional to the brane's tension.  This leads to an automatic cancellation 
of the tension's contribution to the action \cite{cdgv}, so that the four 
dimensional expansion rate is not obviously dependent on the vacuum energy of 
any field theory residing on the brane.  

However, in contrast with codimension-one branes, it is not obvious that one 
can put matter with an equation of state different from pure tension on a 
codimension-two brane \cite{cdgv,Gregory2}.  The reason for this is that in 
order to do so, one has to drop the requirement that the metric be regular at 
the position of the brane.  In order to circumvent this difficulty, Bostock 
{\it et.al.}\cite{Gregory2} have suggested adding Gauss-Bonnet terms to the 
bulk gravitational action.  We have chosen to take a different approach, 
accepting that the singularity will be present, as happens in any theory 
featuring point sources, but dealing with it by constructing a thick brane 
model whose zero thickness limit we will eventually want to study \cite{vc}.  

We will be working with a particular codimension-two braneworld where the 
bulk matter content consists of a cosmological constant and a two-form.  
The interplay between the two compactifies and stabilises the internal 
space \cite{Carroll:2001ih}, which is essentially spherical.  However, 
the presence of branes at the poles will induce a deficit angle, so that 
the internal space will effectively look like an american football, or 
rugby ball.  The relation between the brane tension, denoted by $\sigma^{(4)}$, 
and the deficit angle is given by 
\beqa
\Delta = 8\pi G_6 \sigma^{(4)}.
\eeqa

It can be shown that such a construction has solutions where the four 
dimensional space is flat, dS, or AdS, depending on bulk parameters.  
Indeed, with the two-form given by $F_{ab} = \beta\sqrt{|g_2|}\epsilon_{ab}$ 
where $g_2$ is the internal space metric, and $\epsilon_{ab}$ the two dimensional 
antisymmetric tensor, one finds the following relation between the four 
dimensional expansion rate, $H$, and bulk parameters
\beqa
H^2 = \frac{4}{3}\pi G_6 \left(\Lambda_6-\frac{\beta^2}{2}\right).
\eeqa 
If one adds supersymmetry to the picture\cite{Aghababaie:2003wz,Burgess:2004kd}, 
the dilaton equation of motion forces one to have 
\beqa
\Lambda_6=\frac{\beta^2}{2}
\eeqa
so that the flat brane solutions are actually singled out.  (We will not however 
be including supersymmetry in the construction we consider here).

\section{Regularised branes}

In order to study what happens in this model if we put more general types 
of matter on the brane, we will smooth out the singularity at the brane by 
constructing an equivalent thick brane model\cite{vc}.  Furthermore, we will be 
treating matter as a perturbation to the static ``football'' shaped 
background described in the previous section.  Therefore, our metric 
ansatz will be
\beqa
\label{pertmetric}
ds^2 &=& -e^{2(N_0(r)+N_1(r,t))}dt^2+a_0^2(t)e^{2(A_0(r)+A_1(r,t))}d\vec x^2 
+ (1+B_1(r,t))^2 dr^2\nonumber\\
&&+e^{2(C_0(r)+C_1(r))}d\theta^2+2E_1(r,t)\,dr\,dt.
\eeqa
The bulk action can be written as
\beqa
S_{bulk} = \int d^6x\sqrt{-g}\left(\frac{{R}}{16\pi G_6}-\Lambda_6-\frac{1}{4}F^{ab}F_{ab}\right)
\eeqa
We assume that the only nonvanishing component of the vector 
potential is $A_{\theta}(r,t)=A_{\theta}^{(0)}(r)+A_{\theta}^{(1)}(r,t)$.  
For the sake of generality, we include a possible perturbation of the 6D cosmological constant, 
$\Lambda_6 \rightarrow \Lambda_6+\delta\Lambda_6$.  
The full stress-energy tensor is taken to be of the form
\beqa
T^a_b(r,t)=t^a_b(r,t)+\theta(r_0(t)-r)S^a_b(r,t)+\theta(r-r_*(t)){S_*}^a_b(r,t)
\eeqa
where $t^a_b$ refers to the bulk content, while $S^a_b$ is the core stress 
energy, 
given by
\beqa
S^t_t &=& -\sigma-\rho;\qquad S^x_x = -\sigma+p;\qquad
S^r_r = 0+p^r_r;\nonumber\\
S^{\theta}_{\theta} &=& 0 + p^{\theta}_{\theta};\qquad
S^r_t = 0+p^r_t;\qquad
S^t_r = 0+p^t_r;\nonumber\\
{S_*}^t_t &=& -\sigma-\rho_*;\qquad{S_*}^x_x = -\sigma+p_*\qquad
{S_*}^r_r = 0+{p_*}^r_r;\nonumber\\
{S_*}^{\theta}_{\theta} &=& 0 + {p_*}^{\theta}_{\theta};\qquad
{S_*}^r_t = 0+{p_*}^r_t;\qquad
{S_*}^t_r = 0+{p_*}^t_r.
\label{coreS}
\eeqa 
and we treat the time dependence of the thickness as a perturbation, 
so that $r_0(t) = r_0+\Delta r_0(t)$, $r_*(t) = r_*+\Delta r_*(t)$ and 
\beqa
\theta(r_0(t)-r) &=& \theta(r_0-r)+\delta(r_0-r)\Delta r_0(t) +{O}(\Delta r_0^2) \\
\theta(r-r_*(t)) &=& \theta(r-r_*)-\delta(r-r_*)\Delta r_*(t) +{O}(\Delta r_*^2) .
\eeqa
so that effectively, we can write the stress-energy tensor as
\beqa
t^a_b+s^a_b+{s_*}^a_b
\eeqa
with, e.g.,
\beqa
s^t_t &=& -\sigma\theta(r_0-r) + \left[-\rho\theta(r_0-r)-\sigma\delta(r_0-r)\Delta r_0(t)\right]\\
s^i_i &=& -\sigma\theta(r_0-r) + \left[p\theta(r_0-r)-\sigma\delta(r_0-r)\Delta r_0(t)\right]
\eeqa
and similarly for all other terms.  

Here $\sigma$ represents the tension of the regularised brane, and
$\rho,p$ represent contributions from ordinary matter on the 
standard-model (SM) brane, while starred quantities refer to matter on a hidden
brane which is antipodal to the SM brane on the two-sphere bulk.

The subscripts on the metric and gauge field perturbations indicate their 
order in a perturbative series in powers of $\rho$.  We will furthermore 
assume that time derivatives of the perturbations are of 
${O}(\rho^{3/2})$, which is implied by the usual law for conservation 
of energy $\dot \rho \sim (\dot a/a) \rho \sim \rho^{3/2}$.  

\section{Generalised cosmology}

Solving the system we have just described to linear order in the perturbations, 
one finds that the Friedmann equations emerge through the imposition of boundary 
conditions.  (See \cite{vc} for details).  One further has to be careful to express 
all results in terms of effective four dimensional quantities that are relevant to 
observers confined to the brane rather than the six dimensional quantities that were 
defined above.  This is done by integrating the 6D quantities over the thickness of the 
brane,
\beqa
{S^{(4)}}^a_b = 2\pi\int_0^{r_0} dr\, \sqrt{|g_2|}\, {S^{(6)}}^a_b
\eeqa
which perturbatively leads to
\beqa
\sigma^{(4)}+\rho^{(4)} &=& 2\pi\int_0^{r_0} dr\, e^{C_0}(1+B_1+C_1)(-s^t_t)\\
-\sigma^{(4)} + p^{(4)} &=& 2\pi\int_0^{r_0} dr\, e^{C_0}(1+B_1+C_1)(s^i_i)
\eeqa
and similarly for the other brane.

Also, the 4D Newton constant is related to the 6D one by dimensional reduction,
\beqa
G_6 &=& G_4\times V\\
&=& G_4 \int_0^{2\pi}d\theta\int_0^{\pi/k-\phi}e^{C_0(r)}\\
&=& G_4 \times \frac{4\pi}{k^2\bar k^2}(\bar k^2 +(k^2-\bar k^2)\cos(kr_0))
\eeqa
where we neglect corrections of ${O}(\rho)$.

Expressing the Friedmann equations in terms of these effective 
four dimensional quantities leads to 
\beqa
\label{fried1}
\left(\frac{\dot a_0}{a_0}\right)^2 &=& 
\frac{8\pi G_4}{3}\left(\rho^{(4)}+\rho^{(4)}_*
+\Lambda_{\rm eff}\right)\\
\label{fried2}
\frac{\ddot a_0}{a_0}&=&\left(\frac{\dot a_0}{a_0}\right)^2  
-4\pi G_4\left(\rho^{(4)}+p^{(4)}+\rho^{(4)}_*
+p^{(4)}_*\right)\\
\label{cons1}
\dot\rho^{(4)} &=& -3\frac{\dot a_0}{a_0}\left(\rho^{(4)}+p^{(4)}\right)\\
\label{cons2}
\dot\rho^{(4)}_* &=& -3\frac{\dot a_0}{a_0}\left(\rho^{(4)}_*+p^{(4)}_*\right).
\eeqa
The appearance of the constant $\Lambda_{eff}$ simply reflects the fact that 
our choice to expand around a static background solution was arbitrary, and 
we could just as well have expanded around one of the (A)dS solutions instead.  
The important point is that these are simply the standard Friedmann equations, 
with the possible added contribution from matter on a hidden brane, which 
shows that we do indeed recover standard cosmology from codimension-two braneworlds.  

\section{Generalised deficit angle}

We must now consider how the deficit angle should be defined in 
the case we are considering, where the brane is thick.  From the bulk point of 
view, the radial distance from the brane at $r=R$ is $R-r_0$.  The circumference 
of a circle of radius $R$ is  $2\pi g_{\theta\theta}(R,t)$, while the circumference 
of the brane is $2\pi g_{\theta\theta}(r_0,t)$.  If there is no matter on the brane, so that 
the internal space is perfectly spherical, we would expect that as 
$r_0\rightarrow 0$ and $R\rightarrow 0$, 
\beqa
2\pi[g_{\theta\theta}(R,t)-g_{\theta\theta}(r_0,t)] = 2\pi(R-r_0).
\eeqa
On the other hand, if there is matter on the brane, it will modify 
the previous relation to read
\beqa
2\pi(g_{\theta\theta}(R,t)-g_{\theta\theta}(r_0,t)) = 2\pi(R-r_0)\left(1-\frac{\Delta}{2\pi}\right).
\eeqa
Thus we can define the deficit angle as 
\beqa
\Delta \equiv 2\pi\lim_{R\rightarrow 0}\left[\lim_{r_0\rightarrow 0} 1-\frac{g_{\theta\theta}(R,t)-g_{\theta\theta}(r_0,t)}{R-r_0}\right].
\eeqa
Plugging in the results one gets from solving the linearised equations 
of motion\cite{vc}, we find the following generalised expression for 
the deficit angle
\beqa
\Delta = 2\pi G_6\left(4\sigma^{(4)}+\rho^{(4)}-3p^{(4)}\right)
\eeqa
and similarly around the other brane.  

\section{Discussion}

The first point we wish to emphasise is that the apparent obstruction to 
having arbitrary types of matter on a codimension-two brane is seen to 
stem from the unreasonable expectation that the metric should be regular 
at the position of the brane.  Once this assumption is dropped, there 
is no such obstruction, and it is furthermore possible to smooth out the 
singularities in the metric by considering thick branes.  

Our results show that the answers to our original questions are 
\begin{itemize}
\item codimension-two braneworlds {\it do} lead to FRW cosmology;
\item the deficit angle will respond dynamically to a change in the 
brane tension.
\end{itemize}

Unfortunately, our results also show that there can not be a self-tuning 
mechanism leading to a solution to the cosmological constant 
problem.  This might seem surprising given the fact that the 
cancellation between the deficit angle and brane stress-energy 
which initially led to this hope still holds in the generalised 
model.  However, closer inspection shows that it is the additional 
tuning between the gauge field strength and bulk cosmological 
constant which is spoiled by matter perturbations and leads to 
expansion on the brane.  (See \cite{vc} for a more thourough 
discussion). 

We thus confirm recent work on the subject\cite{Nilles:2003km,Lee:2003wg,Graesser:2004xv} 
which also pointed to the conclusion that in the context of Einstein gravity, codimension-two 
braneworlds did not lead to self-tuning, as can be seen from the absence of massless 
scalars in the low energy effective theory. 

While this conclusion definitely rules out codimension-two braneworlds in Einstein gravity 
as solutions to the cosmological constant problem, the possibility remains open that 
similar supersymmetric models\cite{Aghababaie:2003wz,Burgess:2004kd} might prove more successful
\footnote{In the weeks following the conference where this talk was presented, a preprint came out
\cite{Garriga:2004tq} which makes the claim that supersymmetric models suffer from a fine 
tuning problem, and fall whithin the range of Weinberg's no-go theorem.}.

\section*{Acknowledgments}

The author would like to thank Jim Cline, who collaborated 
on the work this talk is based on, as well as Yashar Aghababaie 
and Cliff Burgess for stimulating discussions, and Canada's 
NSERC for financial support.


\begin{thebibliography}{999}

\bibitem{SN1a}
S.~Perlmutter {\it et al.},
{\it Nature} {\bf 391}, 51 (1998);\\
A.~Riess {\it et al.},
{\it Astron. J.}, {\bf 116}, 1009, (1998).

\bibitem{WMAP}
D.~N.~Spergel {\it et al.},
{\it Astrophys.\ J.\ Suppl.\  }{\bf 148}, 175 (2003)
[arXiv:astro-ph/0302209].


\bibitem{Rubakov:1983bz}
V.~A.~Rubakov and M.~E.~Shaposhnikov,
{\it Phys.\ Lett.\ B} {\bf 125}, 139 (1983).

\bibitem{Randjbar-Daemi:1985wg}
S.~Randjbar-Daemi and C.~Wetterich,
{\it Phys.\ Lett.\ B} {\bf 166}, 65 (1986).

\bibitem{Gibbons:1986wg}
G.~W.~Gibbons and D.~L.~Wiltshire,
{\it Nucl.\ Phys.\ B} {\bf 287}, 717 (1987)
[arXiv:hep-th/0109093].

\bibitem{ADD} 
N.~Arkani-Hamed, S.~Dimopoulos and G.~Dvali,
{\it Phys.\ Lett.\ B} {\bf 429}, 263 (1998)
[hep-ph/9803315],
I.~Antoniadis, N.~Arkani-Hamed, S.~Dimopoulos and G.~Dvali,
{\it Phys.\ Lett.\ B} {\bf 436}, 257 (1998)
[hep-ph/9804398],
N.~Arkani-Hamed, S.~Dimopoulos and G.~Dvali,
{\it Phys.\ Rev.\ D} {\bf 59}, 086004 (1999)
[hep-ph/9807344].

\bibitem{RSI}
L.~Randall and R.~Sundrum,
{\it Phys.\ Rev.\ Lett.\ }  {\bf 83}, 3370 (1999)
[arXiv:hep-ph/9905221].

\bibitem{RSII}
L.~Randall and R.~Sundrum,
{\it Phys.\ Rev.\ Lett.\ }  {\bf 83}, 4690 (1999)
[arXiv:hep-th/9906064].

\bibitem{BDL}
P.~Binetruy, C.~Deffayet and D.~Langlois,
{\it Nucl.\ Phys.\ B} {\bf 565}, 269 (2000)
[arXiv:hep-th/9905012].

\bibitem{CGS}
J.~M.~Cline, C.~Grojean and G.~Servant,
{\it Phys.\ Rev.\ Lett.\ }  {\bf 83}, 4245 (1999)
[arXiv:hep-ph/9906523].

\bibitem{Chen:2000at}
J.~W.~Chen, M.~A.~Luty and E.~Ponton,
{\it JHEP} {\bf 0009}, 012 (2000)
[arXiv:hep-th/0003067].

\bibitem{Carroll}
S.~M.~Carroll and M.~M.~Guica,
arXiv:hep-th/0302067.

\bibitem{Navarro1}
I.~Navarro,
{\it JCAP} {\bf 0309}, 004 (2003)
[arXiv:hep-th/0302129].

\bibitem{cdgv}
J.~M.~Cline, J.~Descheneau, M.~Giovannini and J.~Vinet,
{\it JHEP} {\bf 0306}, 048 (2003)
[arXiv:hep-th/0304147].

\bibitem{vc} 
J.~Vinet and J.~M.~Cline,
arXiv: hep-th/0406141


\bibitem{Gregory2}
P.~Bostock, R.~Gregory, I.~Navarro and J.~Santiago,
arXiv:hep-th/0311074.


\bibitem{Carroll:2001ih}
S.~M.~Carroll, J.~Geddes, M.~B.~Hoffman and R.~M.~Wald,
{\it Phys.\ Rev.\ D} {\bf 66}, 024036 (2002)
[arXiv:hep-th/0110149].


\bibitem{Aghababaie:2003wz}
Y.~Aghababaie, C.~P.~Burgess, S.~L.~Parameswaran and F.~Quevedo,
{\it Nucl.\ Phys.\ B} {\bf 680}, 389 (2004)
[arXiv:hep-th/0304256].


\bibitem{Burgess:2004kd}
C.~P.~Burgess,
arXiv:hep-th/0402200.

\bibitem{Nilles:2003km}
H.~P.~Nilles, A.~Papazoglou and G.~Tasinato,
{\it Nucl.\ Phys.\ B} {\bf 677}, 405 (2004)
[arXiv:hep-th/0309042].

\bibitem{Lee:2003wg}
H.~M.~Lee,
arXiv:hep-th/0309050.


\bibitem{Graesser:2004xv}
M.~L.~Graesser, J.~E.~Kile and P.~Wang,
arXiv:hep-th/0403074.

\bibitem{Garriga:2004tq} 
J.~Garriga and M.~Porrati, 
arXiv:hep-th/0406158. 


\end{thebibliography}
\end{document}